\documentclass[10pt,twocolumn]{article}

\usepackage[sc]{mathpazo} 
\usepackage[T1]{fontenc}
\usepackage{float}
\usepackage{subfig}
\usepackage{textcomp}
\usepackage[hmarginratio=1:1,left=20mm,right=20mm,top=20mm,columnsep=20pt]{geometry}

\usepackage{graphicx}
\usepackage{mathptmx}      
\usepackage{flushend}
\usepackage{hyperref}
\usepackage[numbers,sort&compress]{natbib}
\usepackage{authblk}

\usepackage{abstract}

\title{\vspace{5mm}%
	\fontsize{20pt}{10pt}\selectfont
	\textbf{Three dimensional photograph of single electron tracks through a scintillator}
	}	
\author[1]{\large Mykhaylo Filipenko\thanks{mykhaylo.filipenko@physik.uni-erlangen.de}}
\author[2]{\large Timur Ishakov}
\author[1]{\large Patrick Hufschmidt}
\author[3]{\large Gisela Anton}
\author[4]{\large Michael Campbell}
\author[1]{\large Thomas Gleixner}
\author[3]{\large Gerd Leuchs}
\author[4]{\large Timo Tick}
\author[2]{\large John Vallerga}
\author[1]{\large Michael Wagenpfeil}
\author[1]{\large Thilo Michel}

\affil[1]{\small Friedrich-Alexander University of Erlangen-N\"urnberg Erlangen Centre for Astroparticle Physics Erwin-Rommel-Str. 1, 91058 Erlangen}
\affil[2]{\small Max-Planck Institute for the Science of Light, G\"unther-Scharowsky-Stra\ss e 1/Bau 24, 91058 Erlangen, Germany}
\affil[3]{\small Experimental Astrophysics Group, Space Science Laboratory, University of California, Berkeley, CA-94720, USA}
\affil[4]{\small European Organization for Nuclear Research, CERN, CH-1211, Geneve 23, Switzerland}

\date{\today}

\usepackage{titlesec}
\renewcommand\thesection{\Roman{section}}
\titleformat{\section}[block]{\large\scshape\centering}{\thesection.}{1em}{}

\begin{document}

\maketitle

\begin{abstract}


The reconstruction of particle trajectories makes it possible to distinguish between different types of charged particles. In high-energy physics, where trajectories are rather long, large size trackers must be used to achieve sufficient position resolution. However, in low-background experiments tracks are rather short and three dimensional trajectories could only be resolved in time-projection chambers so far. For detectors of large volume and therefore large drift distances, which are inevitable for low-background experiments, this technique is limited by diffusion of charge carriers. In this work we present a "proof-of-principle" experiment for a new method for the three dimensional tracking of charged particles by scintillation light: We used a setup consisting of a scintillator, mirrors, lenses and a novel imaging device (the hybrid photo detector) in order to image two projections of electron tracks through the scintillator. We took data at the T-24 beam-line at DESY with relativistic electrons with a kinetic energy of 5 GeV and from this data successfully reconstructed their three dimensional propagetion path in the scintillator. With our setup we achieved a position resolution of about 28 \textmu m in the best case.

\end{abstract}

\section{Introduction}

Since the beginning of the 20th century tracking detectors have played an outstanding role for discoveries in experimental particle and nuclear physics. After the first cloud chamber was built by Charles Wilson in 1911, it was possible to observe tracks of charged particles and distinguish between electrons and positrons simply by applying a magnetic field and analyzing the trajectories \cite{bubble_chamber_rev}. With technological progress ongoing the cloud chamber was replaced with the bubble chamber \cite{bubble_chamber_rev} and the spark or wire chamber \cite{wire_chamb_rev} which allowed higher event rates, better position resolution and an automated electrical data read-out. This technology was developed further to the gas-filled multi-wire proportional chamber \cite{wire_chamb_rev} and the gas-filled time-projection chamber \cite{tpc_review} which made it possible to reconstruct trajectories in three dimensions. \par
After the advancement in semiconductor technology silicon strip detectors \cite{ssd_review, lhcb} were developed and used for tracking applications. During the past ten years hybrid active pixelated semiconductor detectors are on the rise and expected to achieve a position resolution down to several \textmu m for vertex tracking applications \cite{lhcb_tp, medipix10}.\par
However, track resolving detectors can are valuable not only for high energy physics experiments but also for low-background experiments like the search for neutrinoless double beta decay \cite{dbreview} or direct detection of dark matter \cite{dmreview}. In such applications the main problem is reduction of background which can produce false positive events. Tracking might be a valuable tool to identify such events and sort them out. For example in neutrinoless double beta decay search experiments, the topological structure of the trajectory of the two decay electrons is different from the trajectory of an electron from a usual beta decay \cite{tp_pub, ahep, next}.\par
Low cross-section experiments require large sensitive detector volumes since the expected event rates are very low. In almost all modern experiments tracking is performed by the read-out of the ionization signal. It means that secondary electrons which were released in a gas or semiconductor material by the primary particle are drifted in an electric field to the collecting electrodes. Conductive wires (in TPCs) or structures in form of strips, rings, electrodes, potential minima (silicon micro-strip detectors, drift detectors, active pixel detectors, DEPFET detectors) collect the released electrons and generate a signal by charge collection or the induction of currents or mirror charges. In micro pattern gaseous detectors, the electrons are drifted towards micromegas structures or GEM-foils. The transversal coordinates perpendicular to the drift direction can be measures by using at least two read-out planes (wire planes in TPCs) or segmentation of the collecting structure in pixels. In gaseous detectors, the drift time of the electrons is measured to reconstruct the drift distance (z-coordinate). \par

\begin{figure}[tb]
\centering
\includegraphics[width=\columnwidth]{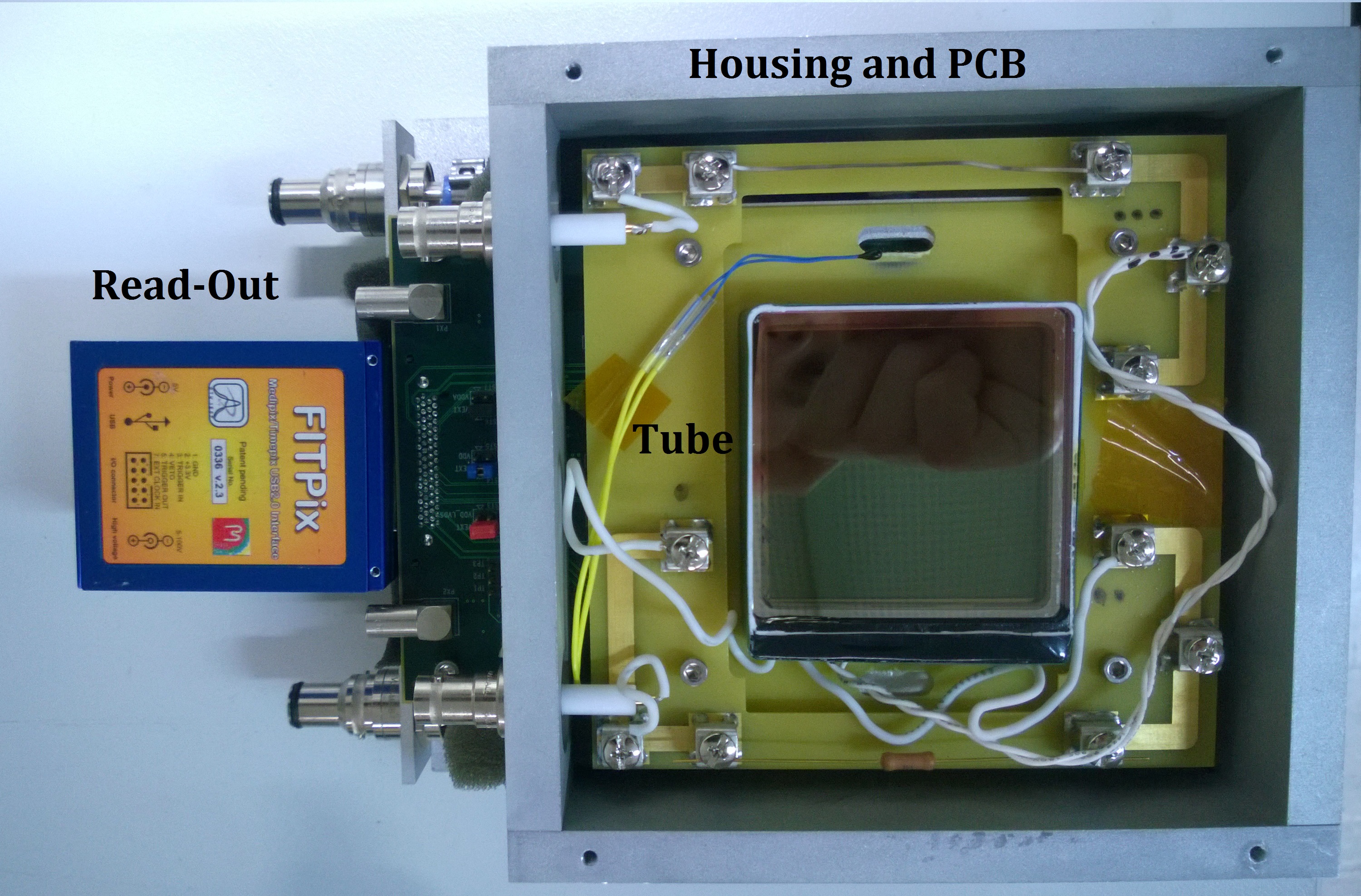}
\caption{A photograph of the hybrid photo detector (HPD) with a FitPix \cite{fitpix} read-out.}
\label{hpd_photo}
\end{figure}

This method has the disadvantage that the position resolution is limited by the diffusion of charge carriers on their way through the sensor material (gas, semiconductor). This limitation makes it difficult to obtain a sufficient position resolution in a detector with large drift distance. For common materials (germanium, cadmium-telluride, liquid xenon) in use for low-background experiments the diffusion radius (the spread of the charge cloud in the plane perpendicular to the drift direction, when it arrives in the read-out plane) is larger then the primary particle's track length after the drift through the sensitive volume which makes it impossible to resolve tracks in a large sensitive volume.\par
This work presents a different approach for the measurement of three-dimensional trajectories of charged particles. The goal is to overcome the diffusion limitation. One possible way to do so is to use the scintillation signal instead of the ionization signal. Scintillation photons are created along the track and do not diffuse on their way through the sensor from the point of origin to the plane where they are detected. One can take advantage of this fact and reconstruct the particle trajectory from multiple two-dimensional projections of the track imaged with a pixelated single photon detector. This publication presents a "`proof-of-principle"' demonstration of this method. We used a plastic scintillator, basic optical components (mirrors and lenses) and a hybrid photon detector which is explained in the next subsection.

\subsection{The Hybrid Photon Detector}

\begin{figure}[tb]
\includegraphics[width=0.95\columnwidth]{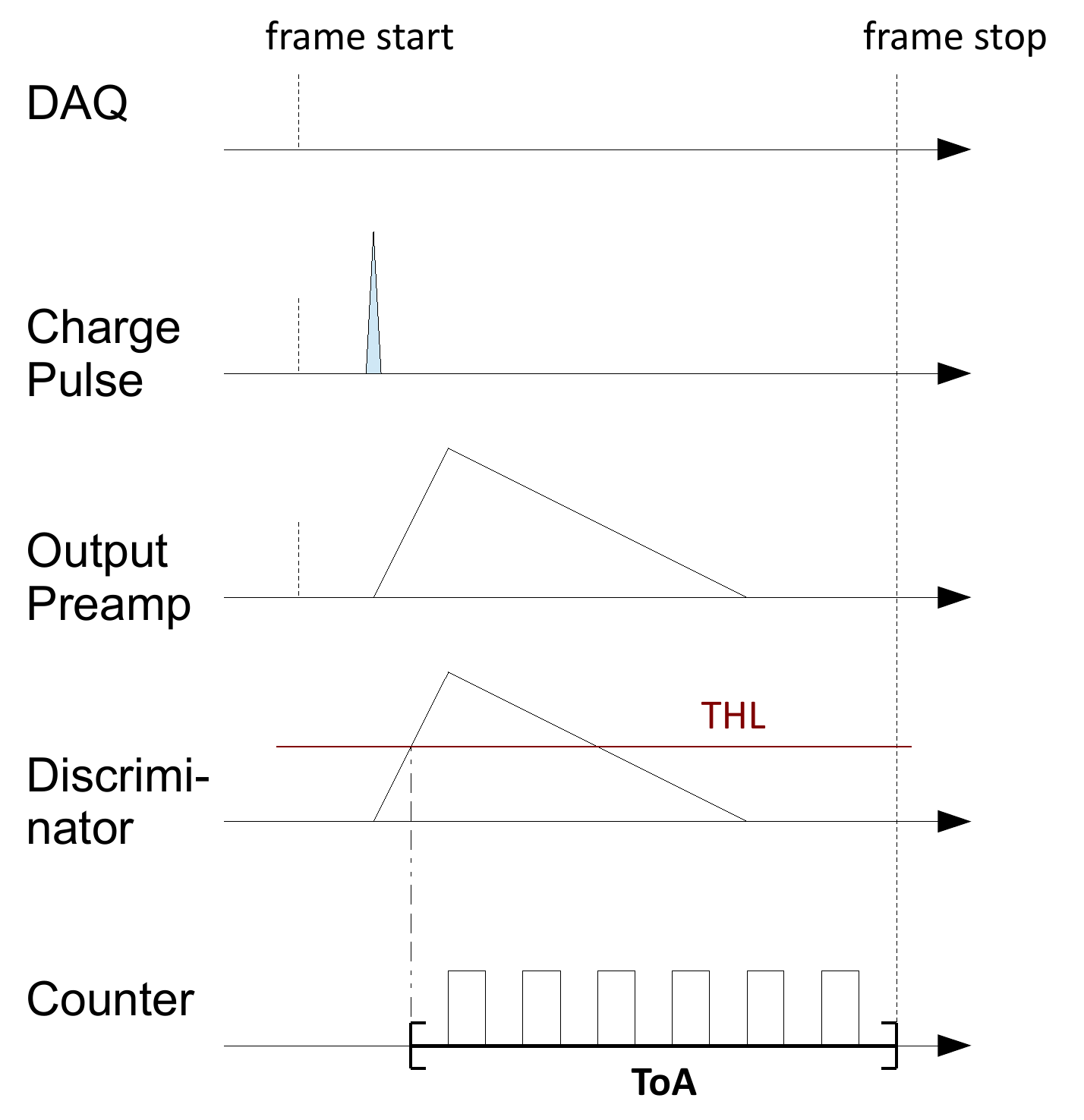}
\caption{The operation of the Timepix in the "time-of-arrival"' (ToA) mode. After the charge pulse is amplified and converted to a voltage pulse, it is discriminated against a threshold level (THL). The time-of-arrival is the number of clock counts between the first time the voltage pulse rises over the THL and the end of the frame.}
\label{timepix_toa}
\end{figure}

The hybrid photon detector (HPD) is a first-of-its-kind detector that combines high spatial and temporal resolution for the detection of single optical photons \cite{hpd_paper}. The detector image vacuum tube consists of a bi-alkali photo-cathode which has its highest quantum efficiency of about 20 \% in the blue/violet spectral range (390 nm). Underneath the photo-cathode there are two microchannel plates (MCPs) in a chevron configuration. Beneath those are four Timepix-ASIC \cite{xavi} arranged in a 2x2 layout (512 x 512 pixels). The tube is sealed under a vacuum pressure of $10^{-10}\textrm{ mbar}$. Photoelectrons which are released from the photo-cathode by optical photons are multiplied in an avalanche-cascade process in the MCPs. The MCP acts as multi-channel photomultiplier. Therefore, one incident photon on the cathode results in an avalanche of about $10^5$ to $10^6$ electrons on the Timepix-ASIC electrodes with the MCPs retaining the information of the input location in the charge cloud centroid. The overall voltage difference between the photo-cathode and the Timepix is 2.4 kV. The Timepix-ASIC is at ground potential whereas the photo-cathode is at -2.4 kV.\par

Because the anode is pixelated, many photons can be detected concurrently. Depending on the mode of operation the timing resolution can be as good as 10 ns and the position resolution as good as 6 \textmu m. The sensitive area is 2.8 cm x 2.8 cm. A photograph of the detector is shown in Fig. \ref{hpd_photo}. Details on the HPD can be found in \cite{hpd_paper}. \par

\begin{figure}[tb]
\centering
\includegraphics[width=\columnwidth]{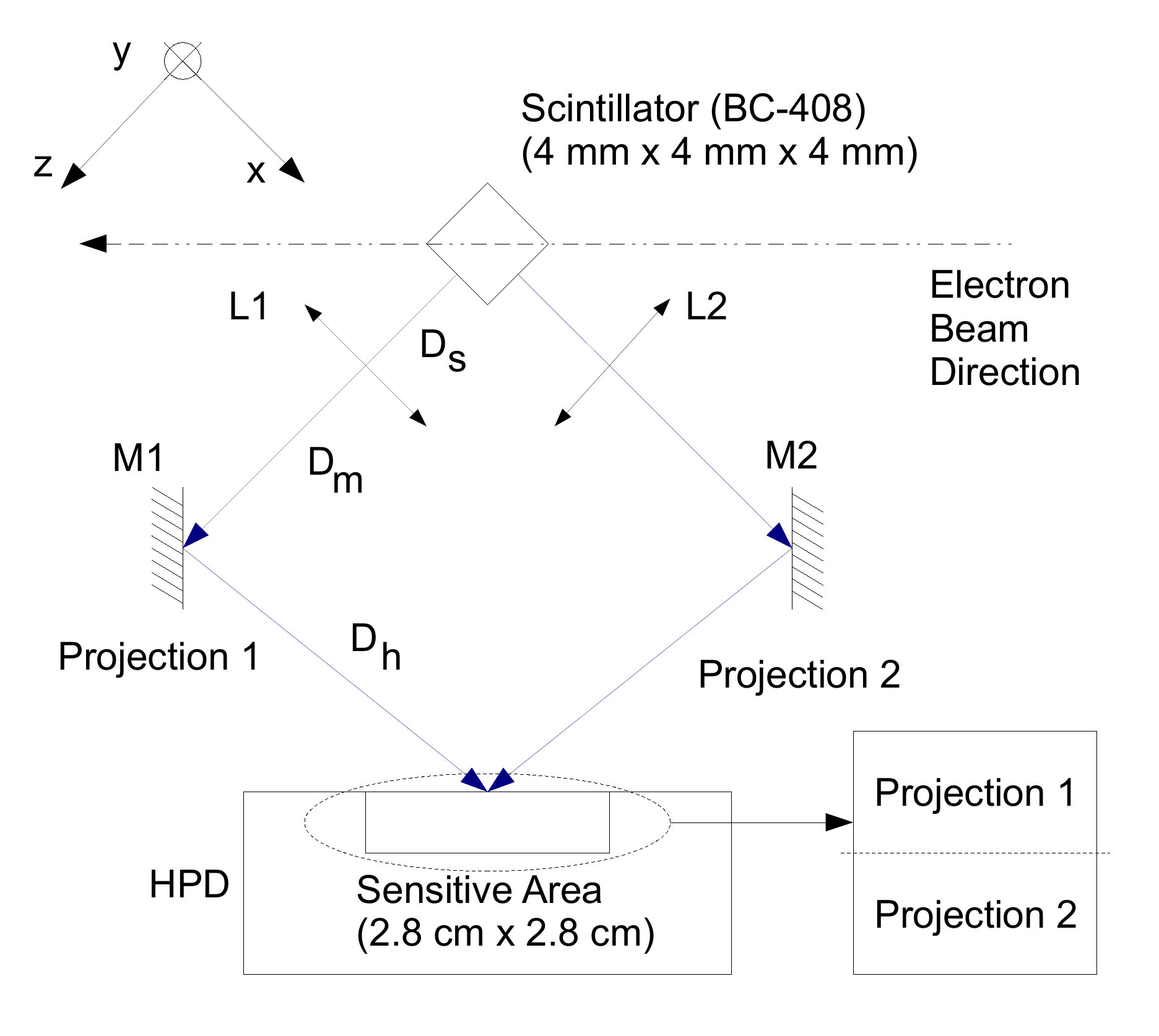}
\caption{Schematic view of the setup used for the measurements. It consists of a plastic scintillator, two lenses (L1, L2) and two mirrors (M1, M2) and the HPD. One projection of the electron track in the scintillator is imaged on the top half of the HPD and the other on the bottom half.}
\label{scheme_setup}
\end{figure}

One Timepix-ASIC has 256 x 256 pixels with a pixel pitch of 55 \textmu m. Each pixel has its own input electrode, connected to an analog circuitry with pre-amplifier and discriminator, and a digital electronics circuitry. In each pixel the charge is collected and converted into a voltage pulse. This is done by a Krummernacher type pre-amplifier. The voltage pulse in each pixel is discriminated against a global threshold which is equivalent to approximately $1000\ e^{-}$. From this point on, the post-processing in each pixel is digital. It operates in three different ways but in this work only the so-called "`time-of-arrival"' (ToA) mode was employed.\par

The ToA mode is illustrated in Fig. \ref{timepix_toa}. When the voltage pulse rises above the threshold a digital register starts counting 
clock cycles until the end of the frame. The number of counted cycles is called "`time-of-arrival"' or ToA. This way, each frame gets a time-stamp wherefore the ToA provides an absolute timing information for every pixel. A frame can have a fixed frame-time or be opened and closed by an external trigger (timing gate). The fastest possible clock frequency is about 100 MHz wherefore every counted clock cycle corresponds to about 10 ns. Details about the Timepix-ASIC can be found in \cite{xavi}.

\section{Experimental Setup}

\begin{figure}
\centering
\includegraphics[width=\columnwidth]{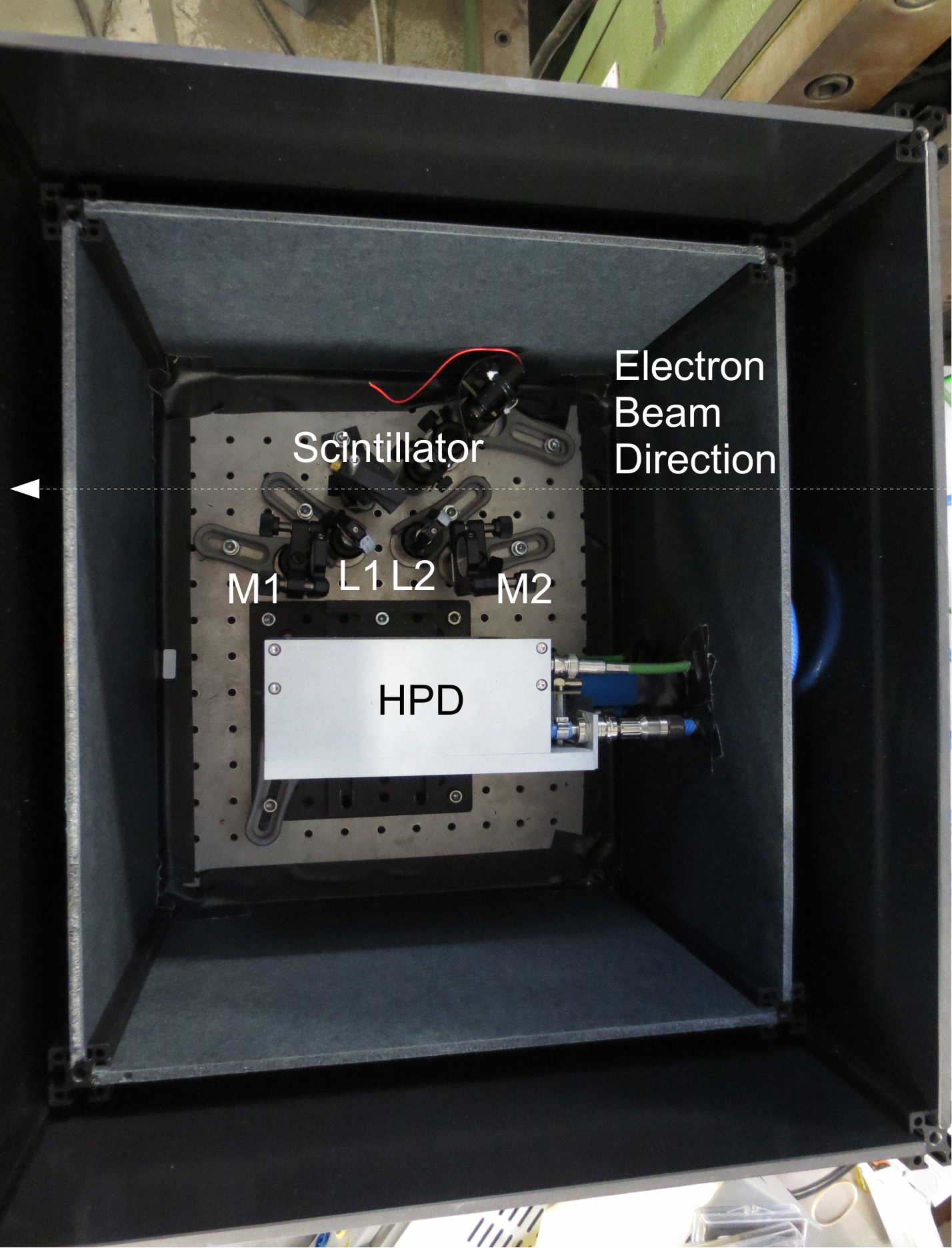}
\caption{A photograph of the setup used for the measurements. The detector and the optics were shielded by two layers (black plastic, dark paper board) from optical photons. The beam direction is shown in the photograph.}
\label{photo_setup}
\end{figure}

\begin{figure}
\centering
\subfloat[]{\includegraphics[width=0.45\columnwidth]{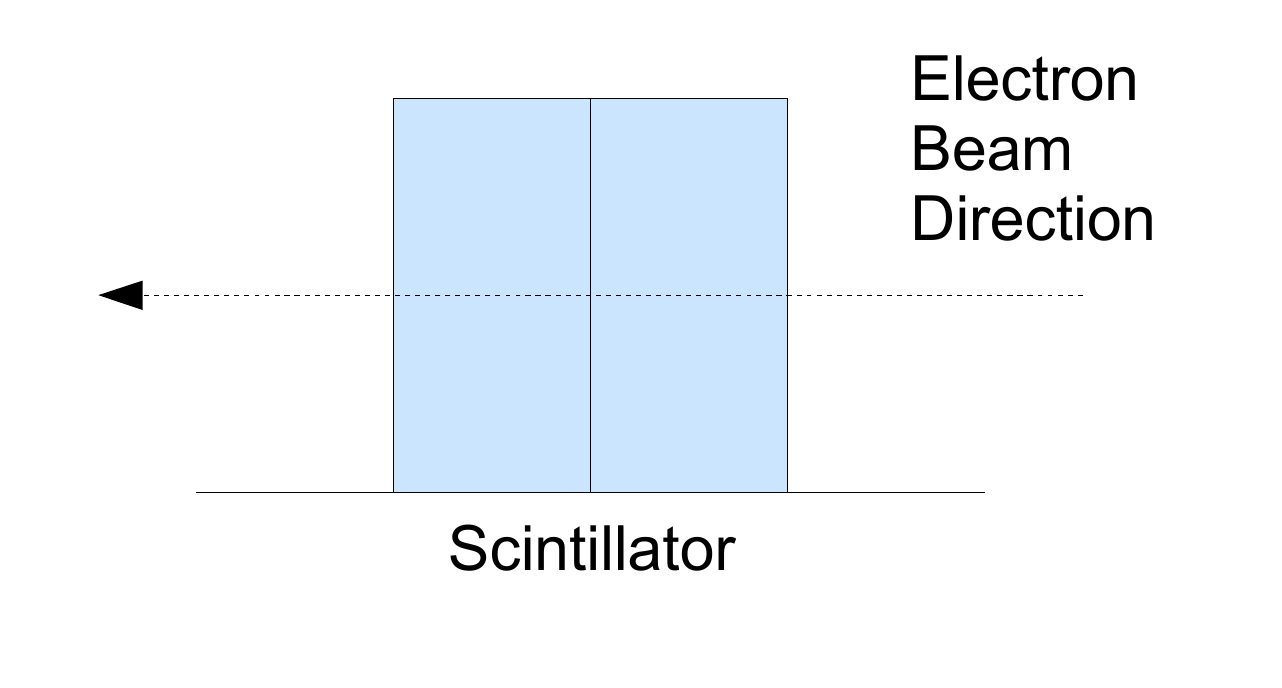}\label{beamd1}}
\subfloat[]{\includegraphics[width=0.45\columnwidth]{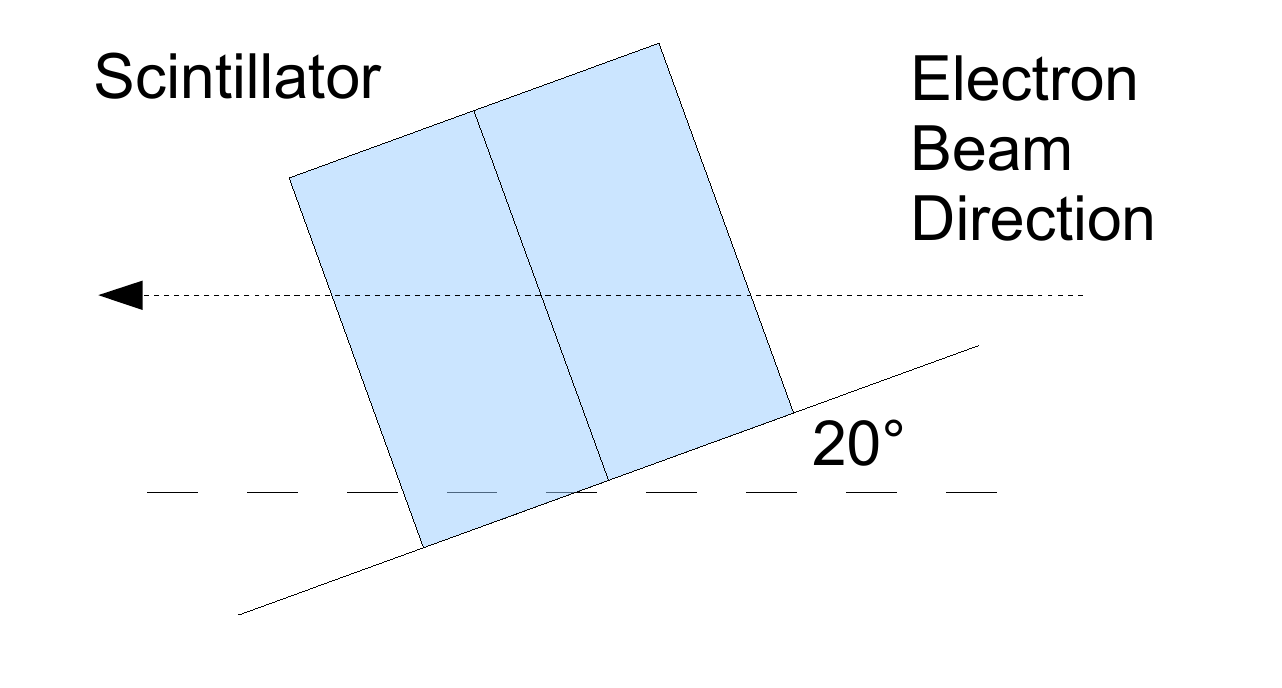}\label{beamd2}}
\caption{The direction of the electron beam through the scintillator.}
\end{figure}

The main idea of this experiment is to reconstruct a three dimensional particle trajectory from multiple two-dimensional projections. When a charged particle propagates through a scintillator, photons are emitted isotropically along the track after relaxation of excited states. An optical system can be used to collect the isotropically emitted light for imaging. In the easiest possible scenario just one single lens and one detector can be used to create a two-dimensional image. If the same scintillator is imaged from multiple perspectives the information can be used to reconstruct a three-dimensional trajectory. \par
For a proof-of-principle demonstration of this method we used a setup as illustrated on Fig. \ref{scheme_setup}. A cubic plastic scintillator (Bicron BC-408) with dimensions of 4 mm x 4 mm x 4 mm was imaged from two orthogonal sides. The scintillation photons have wavelength of about  425 nm. We used two similar Thorlabs LB1761-A bi-convex lenses made of N-BK7 with a focal length of f = 25.4 mm and a diameter of d = 25.4 mm. Two mirrors (Thorlabs BB1-E01) with a size of 25.4 mm in diameter were employed to adjust the optical path. One side of the scintillator was imaged on the top half of the HPD and one side on the bottom half. \par

The distances between the optical elements were $D_s = 34 \textrm{ mm}$, $D_m = 25 \textrm{ mm}$ and $D_h = 77 \textrm{ mm}$. With this geometry we calculated a magnification of 3.3 and a light detection efficiency of about 3 \% per image, taking into account the geometrical acceptance, the losses in the optical elements and the quantum efficiency of the photocathode. Although the scheme suggests, that the angles were 45\textdegree, in the real setup the angles were about 4\textdegree$\ $off. With our geometry, the focal point spread function (or blur circle) $\sigma_f$ at a distance $d_f$ from the focal plane can be estimated as $\sigma_f \approx d_f$. It was calculated as described in \cite{blur_circle}. Therefore, we expect to see sharp tracks only from an inner part of the scintillator around the focal plane (500 \textmu m to each side of the focal plane).  \par
We used two layers to shield the setup from optical photons: The inner layer is a dark paper board from Thorlabs; the outer layer are walls made of 5 mm thick black plastic. A photograph of the setup is shown in Fig. \ref{photo_setup}. The dark rate in the HPD was 2000 counts per second, which corresponds to the intrinsic dark rate of the detector.\par

We used the setup at the DESY testbeam T-24 where the scintillator was penetrated by electrons with a kinetic energy of 5 GeV. Electrons of this energy can be regarded as minimal ionizing and therefore most of the time their trajectories through the scintillator are straight lines. Concerning the stopping power of the plastic scintillator at this energy, we expect about 100 to 150 detected photons per image.\par
The maximum digital counter value in the Timepix ToA mode is 11810 and as we used a clock of 10 MHz (100 ns per clock cycle), the maximum frame integration time was limited to 1 ms. Since the read-out speed of the data acquisition system is limited as well, we could only start a frame every 0.1 s. To increase statistics (the number of measured events), we used the trigger signal from the beam monitor (two crossed scintillator panels with a PMT attached) for the frame-stop signal in order to have a higher probability to see one electron during one frame. The flux density of electrons was about $1000 \frac{e^{-}}{s \cdot cm^2}$.\par

We took data at two different orientations between the scintillator and the beam. In first case the beam penetrated the scintillator from one edge of the scintillator to the other as shown in Fig. \ref{scheme_setup} and Fig. \ref{beamd1}. In the second case the setup was titled by 20\textdegree$\ $with respect to beam axis (Fig. \ref{beamd2}).

\begin{figure}[tb]
\centering
\subfloat[]{\includegraphics[width=0.95\columnwidth]{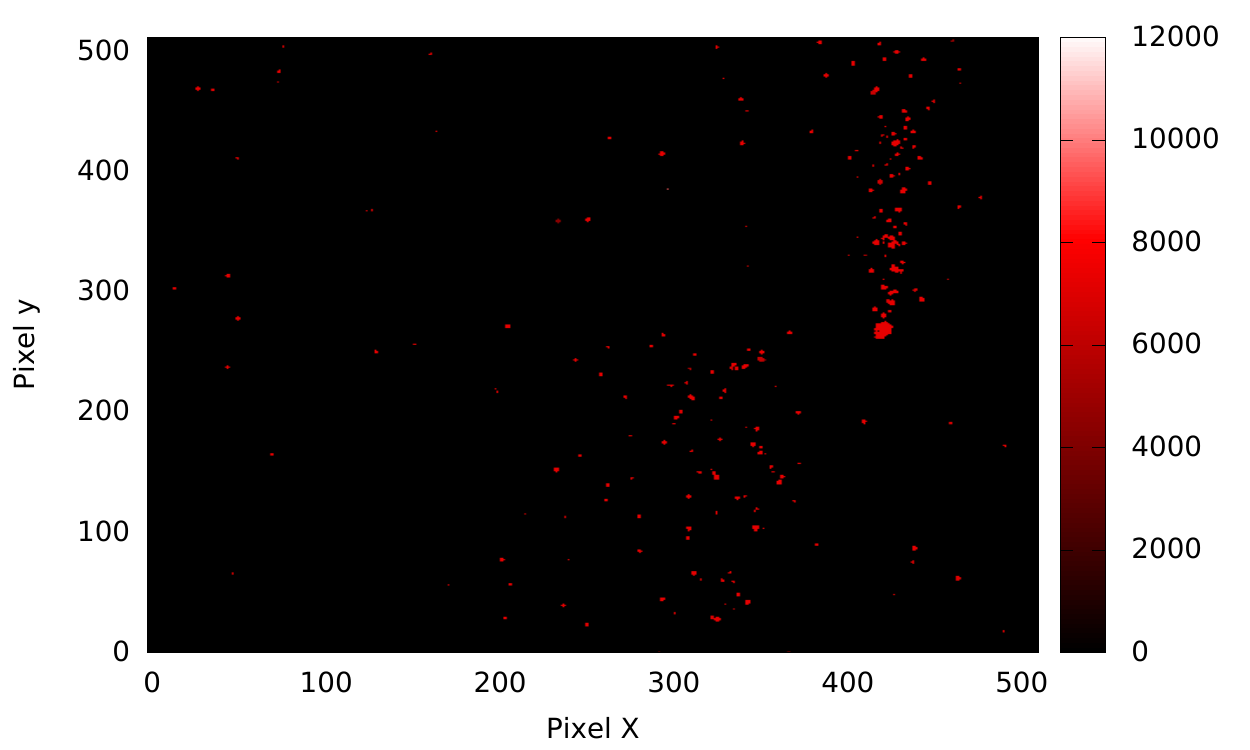}\label{examp_track}}

\subfloat[]{\includegraphics[width=0.95\columnwidth]{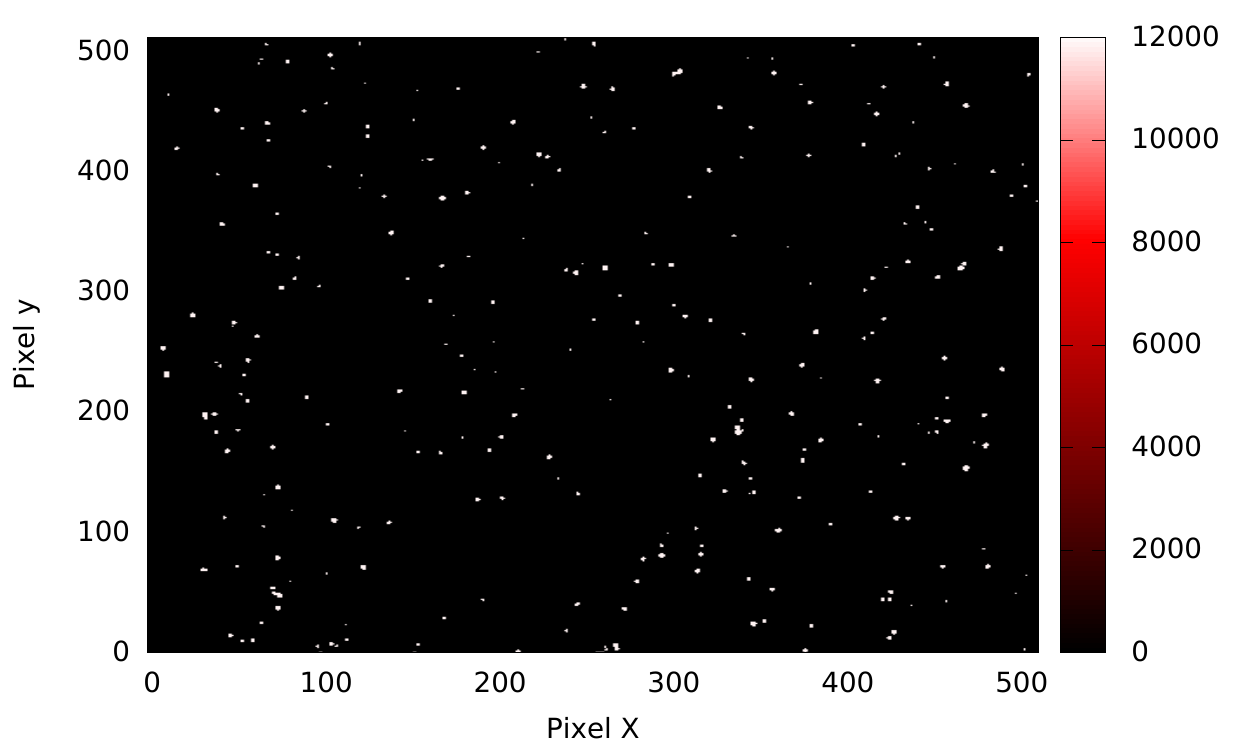}\label{examp_stars}}
\caption{Example of two typical frames: One with a track (a) and one with noise only (b). The color bar indicates the time-of-arrival measured in each triggered pixel.}
\end{figure}

\section{Data Analysis}

\begin{figure}
\centering

\subfloat[]{\includegraphics[width=0.95\columnwidth]{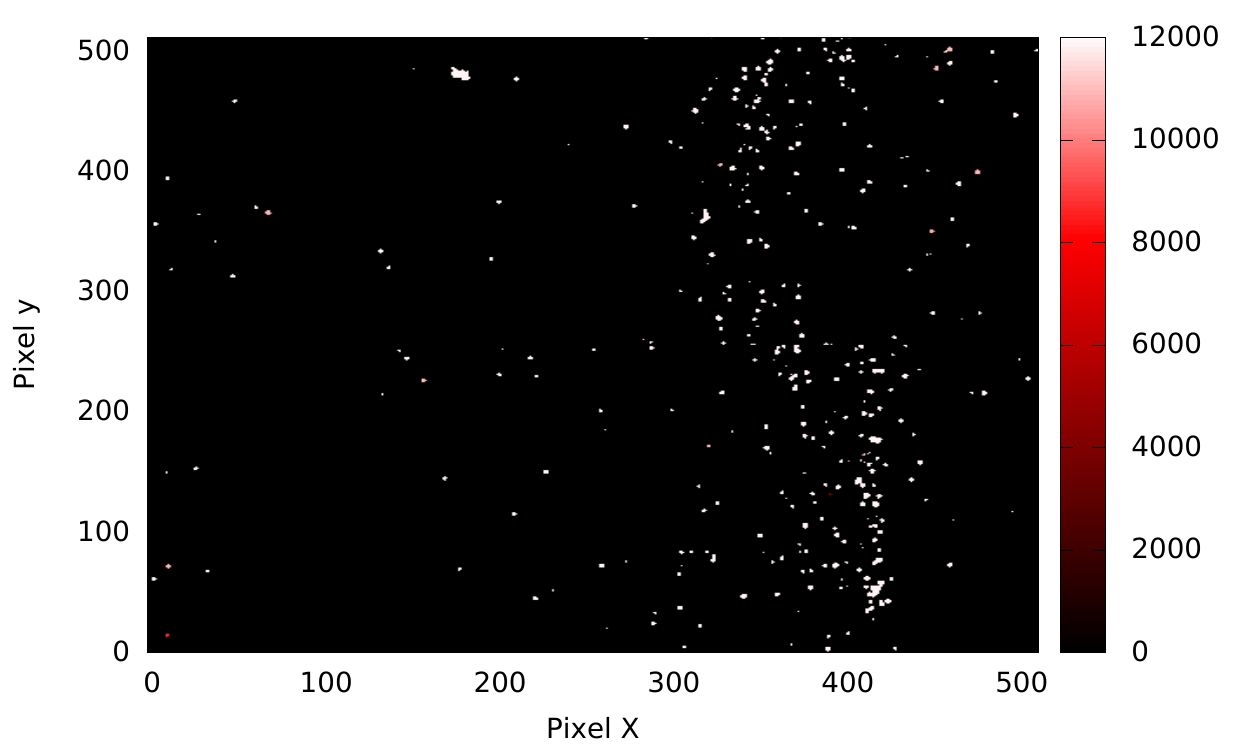}\label{hough_bad_frame}}

\subfloat[]{\includegraphics[width=0.9\columnwidth]{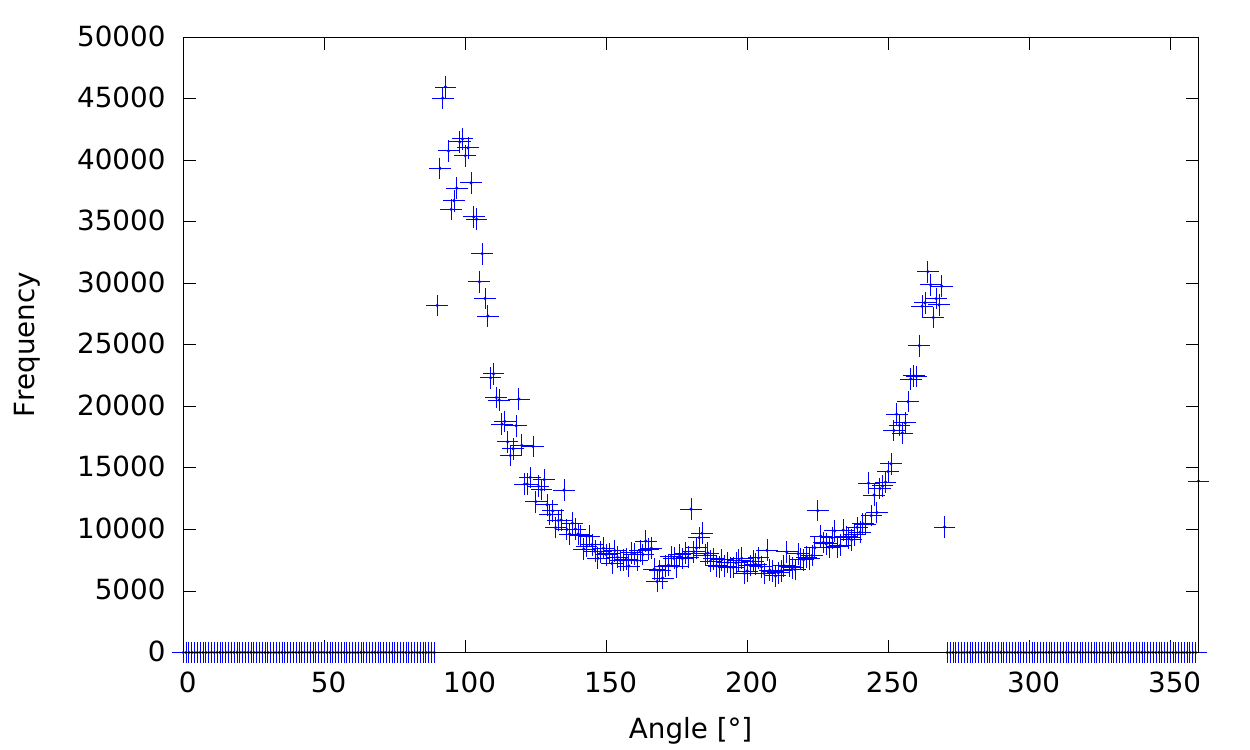}\label{hough_good_spec}}

\subfloat[]{\includegraphics[width=0.95\columnwidth]{images/example_frame_stars.pdf}\label{hough_bad_frame}}

\subfloat[]{\includegraphics[width=0.9\columnwidth]{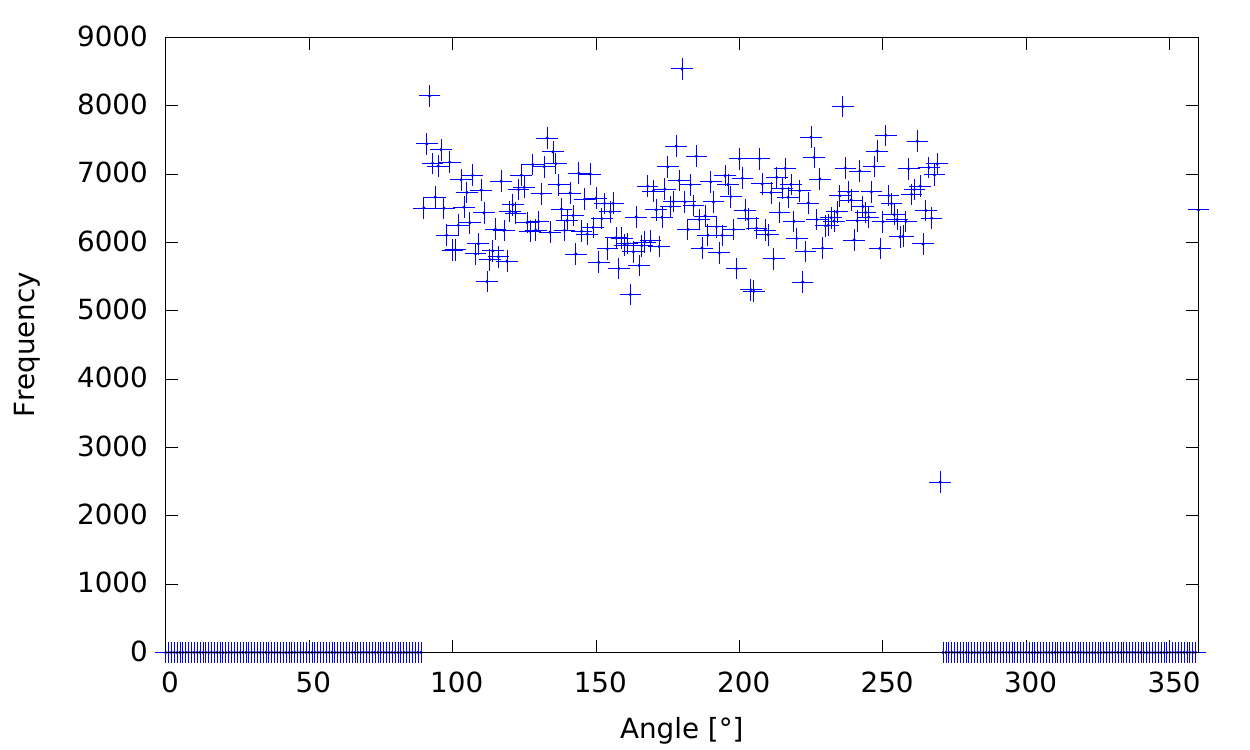}\label{hough_bad_spec}}

\caption{Examples of images ((a) and (c)) and their reduced Hough-transformations ((b) and (d)). If a track is on the frame, peaks which belong to the direction of the track and a clear minimum appear in the transformation. In the opposite case the distribution is flat. The color bar indicates the time-of-arrival measured in each triggered pixel.}
\end{figure}

\begin{figure}
\centering
\subfloat[]{\includegraphics[width=0.95\columnwidth]{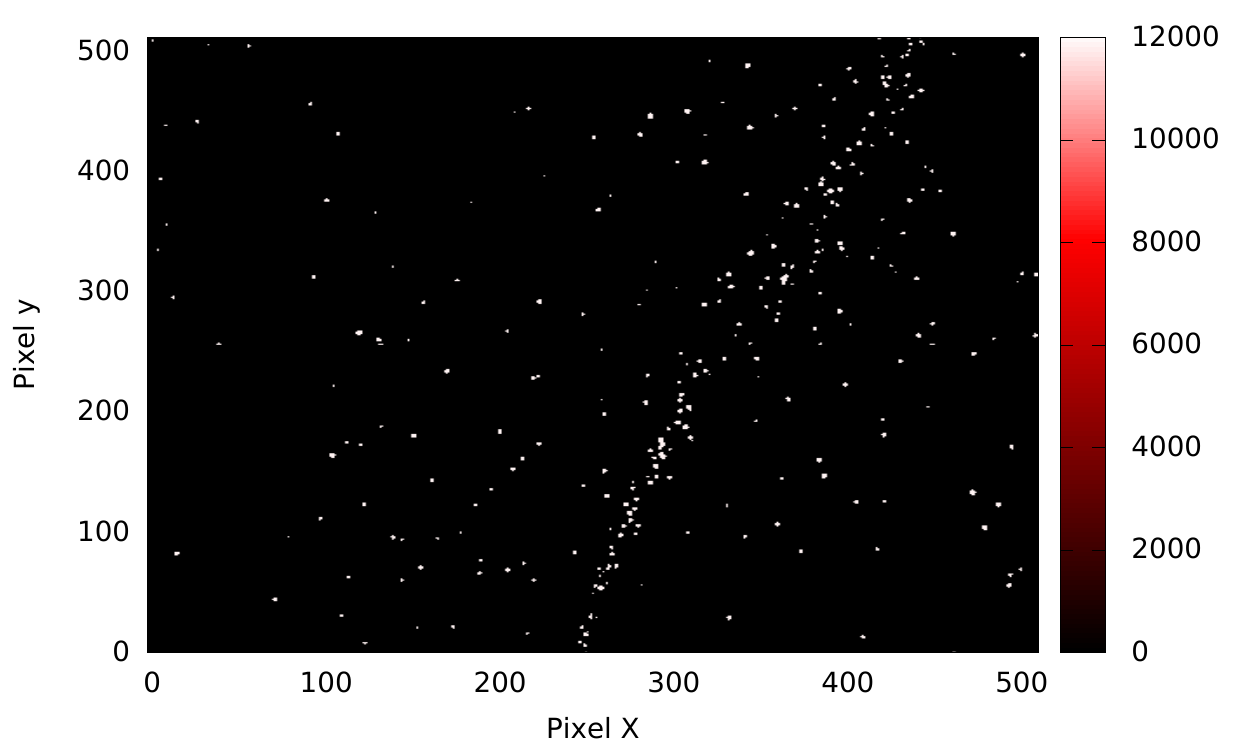}\label{raw_frame}}

\subfloat[]{\includegraphics[width=0.9\columnwidth]{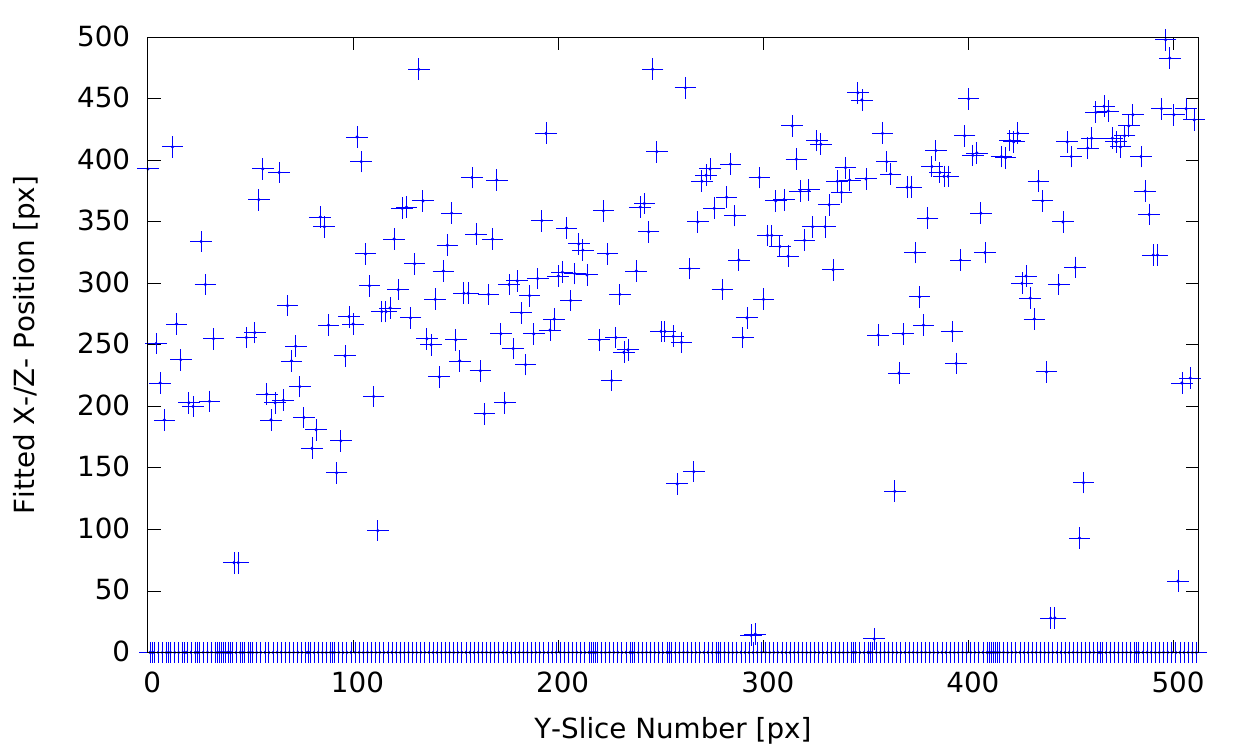}\label{slices_res_raw}}

\subfloat[]{\includegraphics[width=0.95\columnwidth]{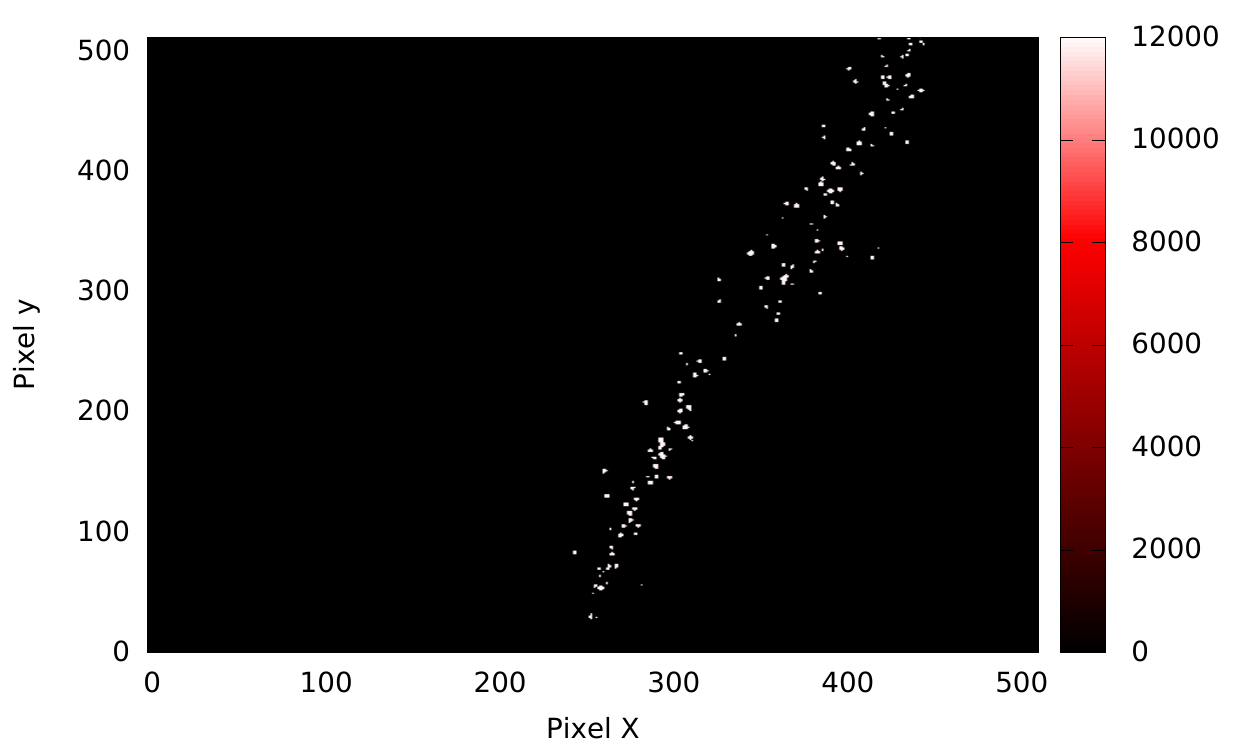}\label{clean_frame}}

\subfloat[]{\includegraphics[width=0.9\columnwidth]{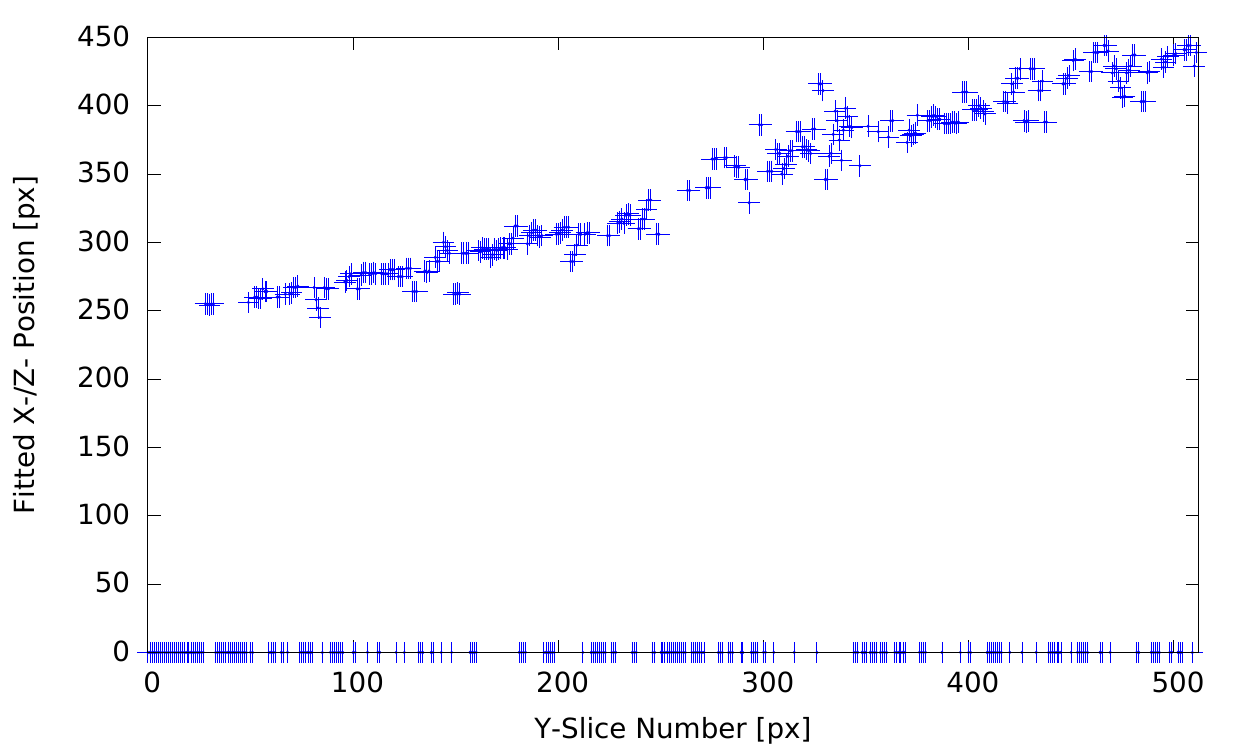}\label{slices_res_clean}}
\caption{Examples of images ((a) and (c)) and the fitted x-/z- position for every y-slice before (b) and after (d) frame cleanup. The color bar indicates the time-of-arrival measured in each triggered pixel.}
\end{figure}

A typical frame contains either a track (as in Fig. \ref{examp_track}) as expected or some random hits which we will call "night sky-frames" (Fig. \ref{examp_stars}). Such frames happen if the electrons path through the scintillator was out of the optical focus, or the trigger gate closed without any electron passing through the scintillator. This was possible since the senstive area of the beam monitor was larger then the cross section area of the scintillator in the setup. \par

\begin{figure}[tb]
\centering
\subfloat[]{\includegraphics[width=\columnwidth]{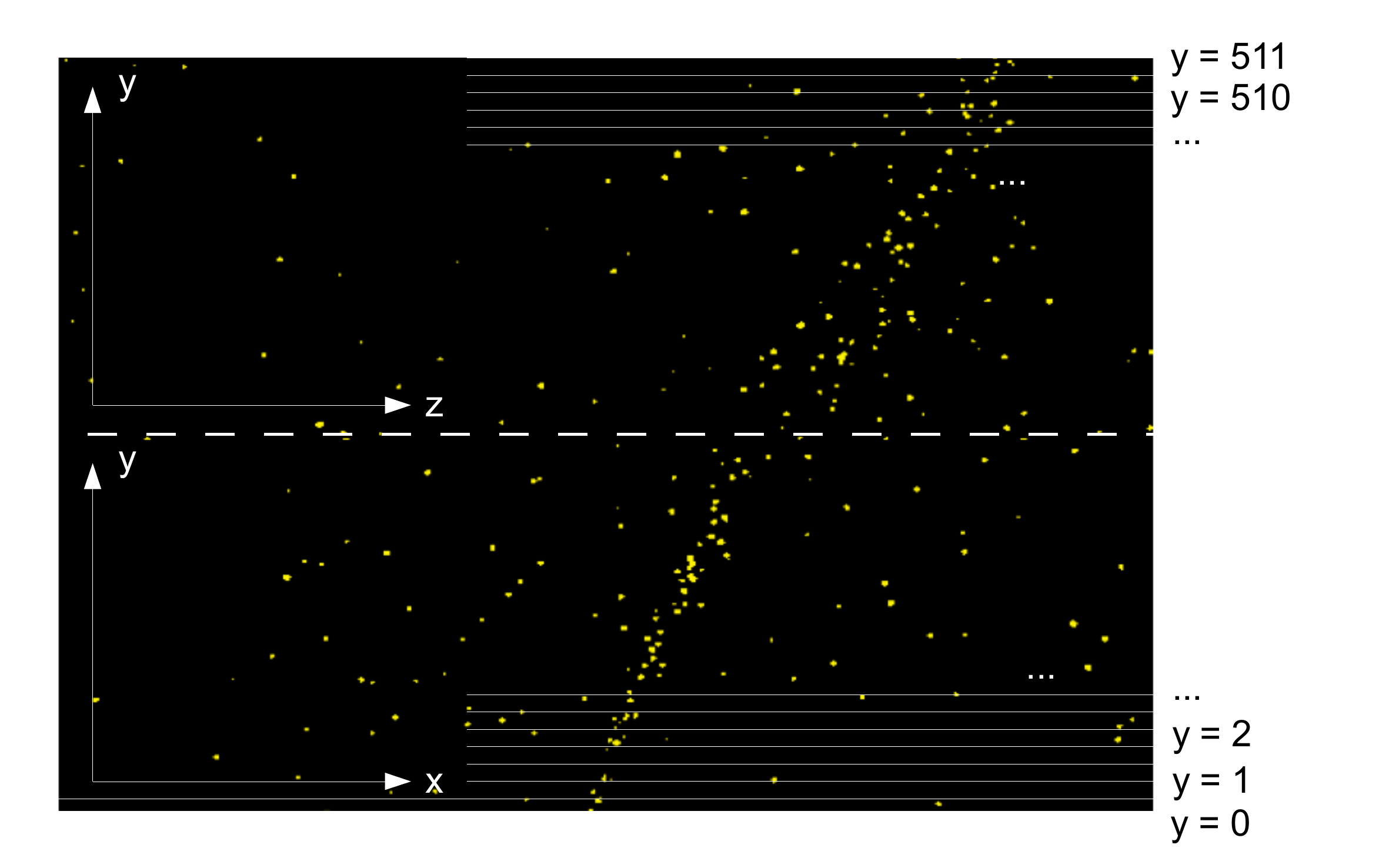}\label{slicing}}

\subfloat[]{\includegraphics[width=\columnwidth]{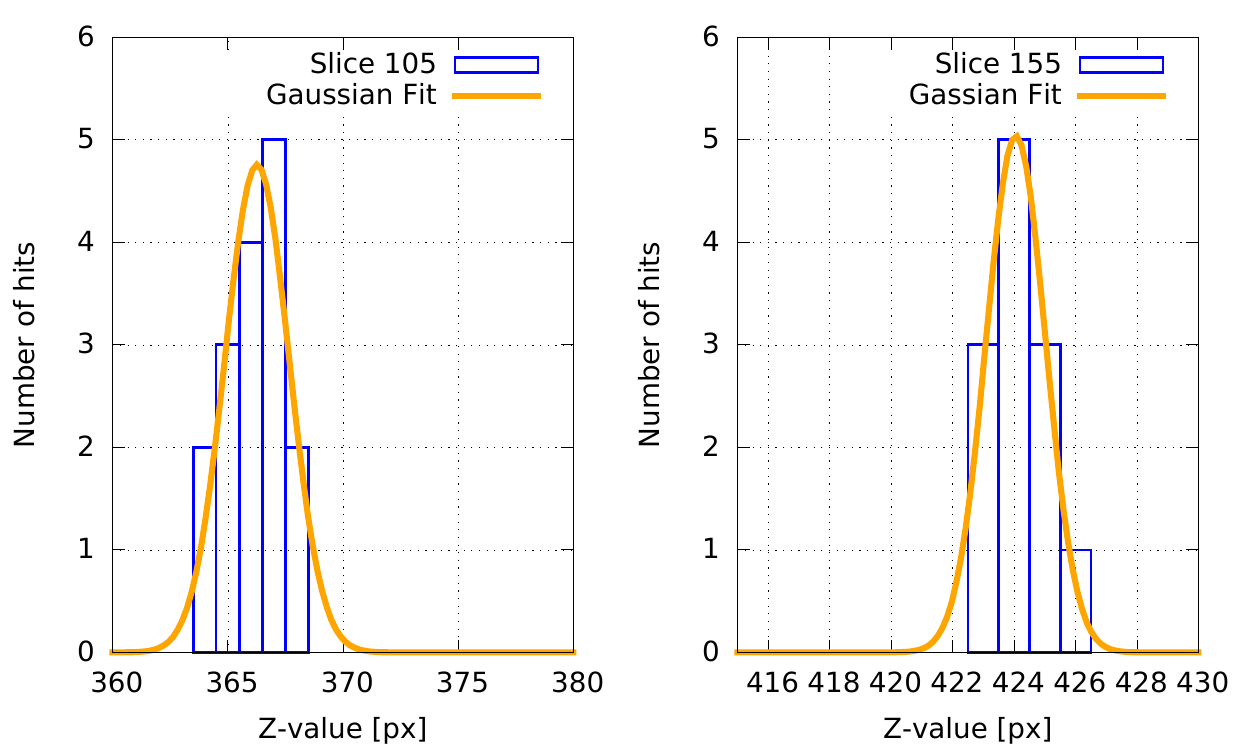}\label{hist_slices}}
\caption{An illustration of the slicing procedure. Each frame is divided into slices parallel to the x-/z-axis (a). In each slice the mean x-/z-position is determined by a fit with a Gaussian. For each slice we obtain a histogram as in (b). The histograms shown here are summed over 3 consecutive slices for better statistics.}
\end{figure}

Consequently, the first step was a frame selection due to the following criteria: First, we removed frames where less then 400 pixels were triggered. About 100 photons were expected per image and due to the electron avalanche of MCP amplification impinging on the Timepix ASIC, one photon usually triggers 3 to 5 pixels. \par

\begin{figure}
\centering
\includegraphics[width=\columnwidth]{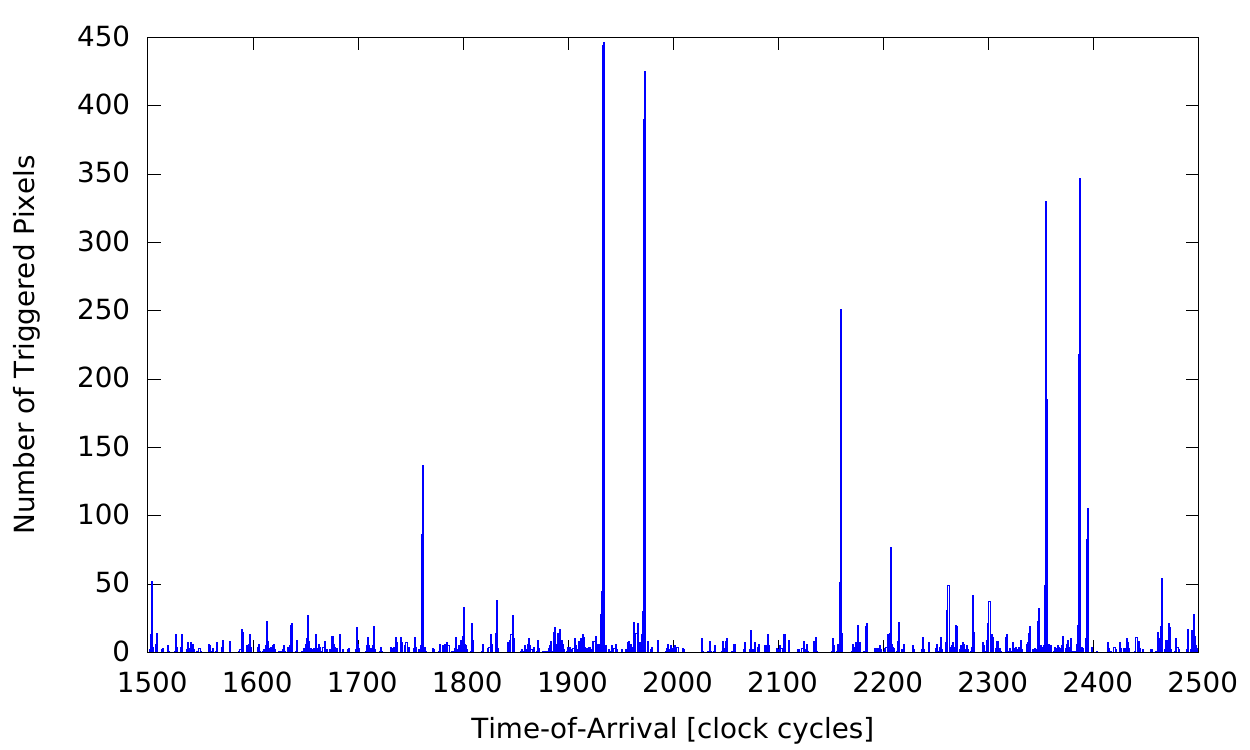}
\caption{Examplary part of the ToA spectrum of a full run. Each peak belongs to the coincident detection of several photons; thus, to one electron track.}
\label{toa_spec}
\end{figure}

Secondly, we performed a reduced Hough-transformation of the frame \cite{hough}, i.e. we calculated the angle between the x-axis and the connecting line of any two triggered pixels in each frame and sorted the angles into a histogram. A typical reduced Hough-transformed for a frame containing a track is shown in Fig. \ref{hough_good_spec} with the corresponding frame; a "night-sky-frames" with the corresponding reduced Hough-transformed is shown in Fig. \ref{hough_bad_spec}. We calculated the mean to the lowest 30 \% of the data points in the reduced Hough transformed and then the absolute deviations of each data point to this mean. If the normalized sum of these deviations was lower then 800 the frame was ignored. \par 

For the three dimensional trajectory reconstruction we "sliced" the two dimensional images along the y-axis, parallel to the x-/z-axis as shown in Fig. \ref{slicing}. We obtain 512 y-slices in total; 256 y-slices for the xy-plane and 256 slices for the xz-plane. The histograms of two single slices are shown in Fig \ref{hist_slices}. The average x-/z- position in every slice is determined by the mean value of a Gaussian fitted to the data in each histogram. The fit is performed with the maximum likelihood method as the number of hits per slice is very low. If the number of hits in the slice is zero, the slice is ignored. If the number of hits in the slice is lower then 4, the average x-position is used instead of the fit since in these cases the fit failed for the most time.\par

The result of this procedure is shown in Fig. \ref{slices_res_raw}. In this plot, the calculated x-value (on the y-axis) is plotted for every y-slice (on the x-axis). One can see a clear trend belonging to the actual track which is in-between deviations due to random hits on the matrix. As discussed earlier such hits can happen due to dark noise or false triggering. Such hits do not belong to the track but affect the reconstruction. Therefore, it was necessary to "clean up" the frame before reconstruction.\par

For this purpose, we calculated the center position for each cluster of adjacent triggered pixels (one photon detection event) and determined the number of neighbouring photon detection events within a chosen radius. After choosing a particular radius, clusters are removed from the frame if the number of neighbours within that radius is below a chosen threshold value. A choice of 30 pixels for the radius and of 8 neighbours for the threshold value produced reasonable and reliable results. One exemplary frame is shown before and after cleanup in Fig. \ref{raw_frame}, \ref{clean_frame} with the y-slices (Fig. \ref{slices_res_raw}, \ref{slices_res_clean}), respectively. We can see that by this procedure the large deviations from the main line are strongly reduced.\par

The reconstruction of the three dimensional trajectories was performed by correlating the y-slices due to their y-value. Every point in the trajectory is referred to as the number of the y-slice, the calculated x-position in that slice in the xy-plane and the calculated z-position in that slice in the zy-plane. To ensure the fact that the photons used for the reconstruction of one track belong to the same track, we used only hits which occur at the same time, i.e. have the same ToA-values within the ToA-resolution. That means, that the ToA values belong to one peak. These can be recognized in the ToA-spectrum of a frame. A typical spectrum for one run in the region of interest is shown in Fig. \ref{toa_spec}. For instance the peaks at about 1930 clock cycles or 1980 cycles belongs to one electron trajectory, respectively.

\section{Results}

\begin{figure}[tb]
\centering
\subfloat[]{\includegraphics[width=0.95\columnwidth]{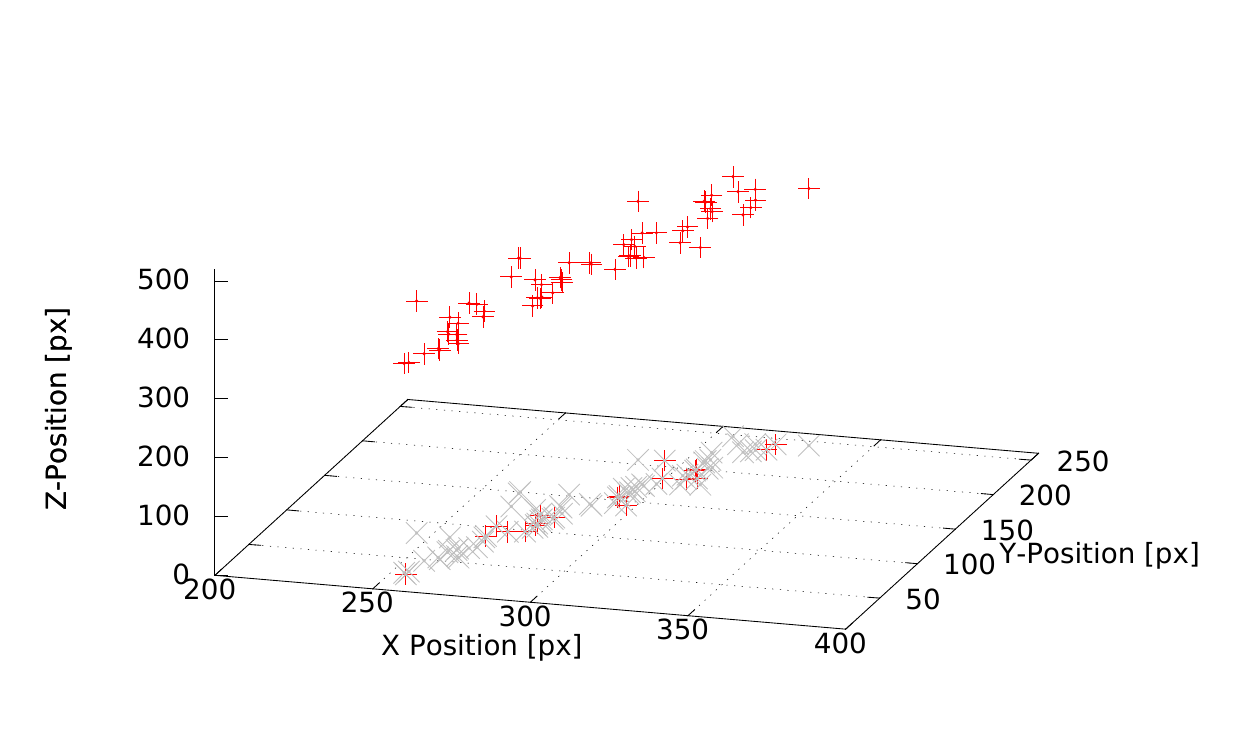}}

\subfloat[]{\includegraphics[width=0.95\columnwidth]{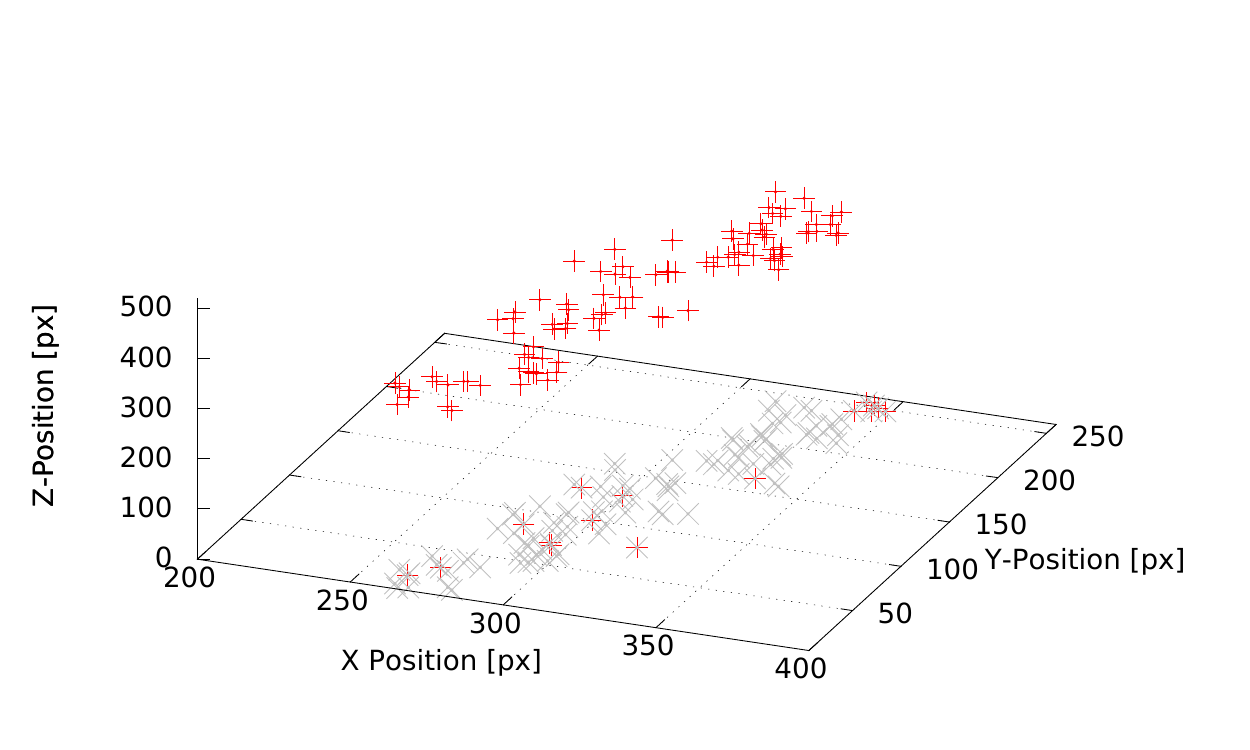}}
\caption{Two examples of reconstructed three dimensional electron trajectories.}
\label{3d_tracks}
\end{figure}

Two typical reconstructed tracks are shown in Fig \ref{3d_tracks}. As expected one can see straight lines as trajectories. The main reason for the deviations from the straight line or "broadening" is due to the fact that electrons propagate not through the focal plane of the imaging system and therefore the trajectory is blurred. Depending on the distance from the focal plane the blurring affects the achievable position resolution.\par

\begin{figure}[tb]
\centering
\includegraphics[width=\columnwidth]{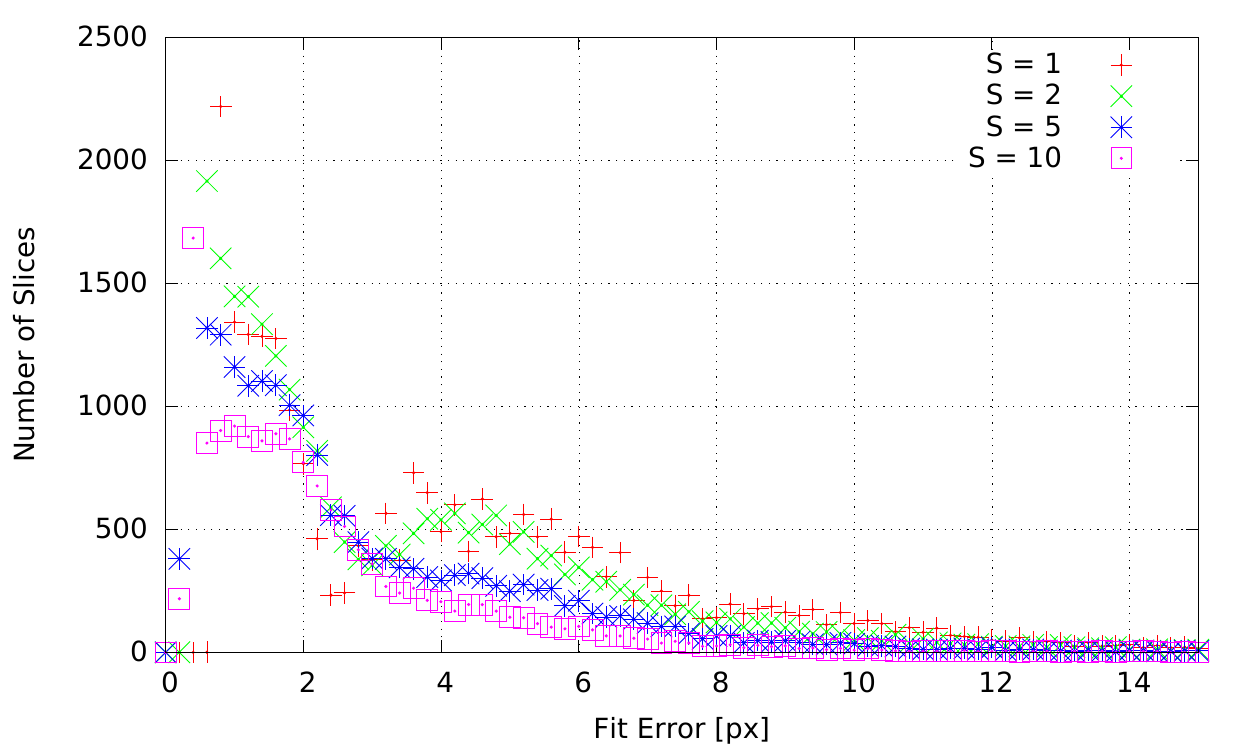}
\caption{The distribution of the fit errors for the slices with different slice binning S for one run.}
\label{res_fiter}
\end{figure}

The position resolution was assessed in two ways: First, the error mean (fit error of the mean) obtained from the Gaussian fits to the slices was regarded as a measure of the position resolution. This is an estimate for how accurate the position of a track through the scintillator can be determined at any point of the image and therefore measures how close two tracks can be next to each other to be distinguishable. The fit error depends on the "slice binning". Slice binning means how many slides are merged together into one histogram for fitting (If the slice binning is 2, we have 256 slices instead of 512; if the slice binning is 4, we have 128 slices etc.). A distribution of the fit errors accumulted over one run (many tracks) is shown in Fig. \ref{res_fiter} for a slice binning of S = 1, 2, 5 and 10. The second peak in the distribution disappears with larger binning. We were not able to interpret this effect by an analysis of our data. As resolution we used the average value (weighted integral divided by the total number of slices) of the distribution multiplied by the pixel size (55 \textmu m) and divided by the magnification (3.3). The results are summarized in table \ref{tab_res}. At this point it is interesting to mention that if only the first peak is considered, the resolution seems to be as good as 28 \textmu m.

\begin{figure}[tb]
\centering
\includegraphics[width=\columnwidth]{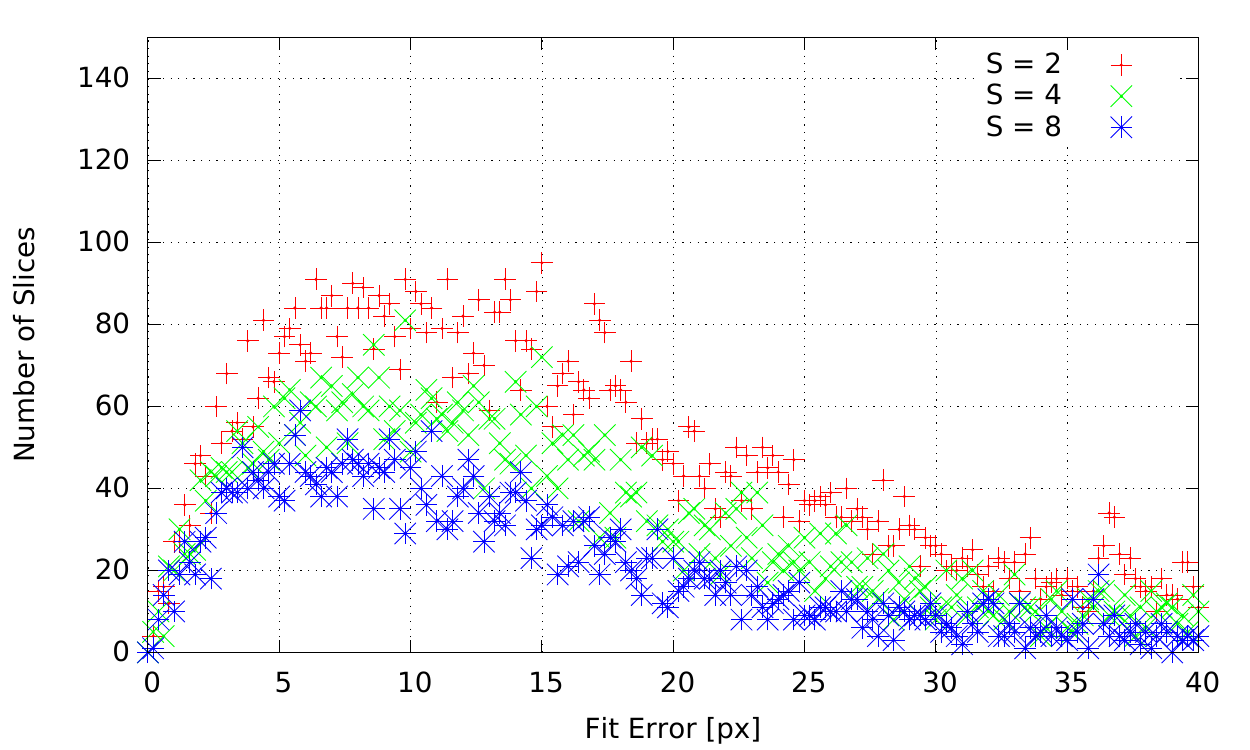}
\caption{The distribution of the fit error for the slices with different slice binning S for one run.}
\label{res_line}
\end{figure}

Secondly, we fitted a straight line to our data and regarded the distribution of residuals (deviations of the data from the fitted line) in each point as the position resolution. As the fitted line has very little fit uncertainties due to the large number of points in the track, it can be regarded as the actual track. Hence, the deviations give us an estimate for how close to the actual track a reconstructed point is and how much the positions of the detected photon scatter around the actual track. In Fig \ref{res_line} the distribution of residuals is shown for different binning taken into one histogram for a whole run (many tracks). Also, in this case we calculated the resolution as the average value (weighted integral divided by the total number of slices) of the distribution multiplied by the pixel size (55 \textmu m) divided by the magnification (3.3). The results are summarized in table \ref{tab_res}. 

The table suggest that thicker slices produce a better resolution in the sense it was defined above. This is true in our particular experiment, since the track is a straight line and large slices allow to fit higher statistics. The result is the reduction of the fit uncertainty. However, in the case of curly tracks this might not be possible, in particular because stronger binning reduces the resolution in the direction perpendicular to the slices. Nonetheless, a resolution of 70 \textmu m is possible with slices of thickness 1 which is an encouraging result for the first test.\par

\section{Conclusions and Outlook}

We have demonstrated that it is possible to reconstruct three dimensional trajectories of particle propagation through a scintillator by imaging two projections of the track on a pixelated single photon detector (the HPD). Already with our simple optical setup, which consisted of two mirrors and two lenses, a remarkable resolution was achieved. With an optical magnification of 3.3 we could achieve a resolution as defined above ranging from 28 \textmu m (in the best cases) to 202.3 \textmu m depending on the method of evaluation.\par
The resolution of the method presented here is mainly limited by the optics used for collecting the scintillation light. However, the main focus was a first "proof-of-principle" demonstration and therefore we concentrated on the easiest possible setup which allowed sufficient light collection efficiency for track imaging, i.e. placing the lense as close to the scintillator as possible. For practical experiments where larger volumes should be imaged a higher depth of focus could be realized with larger lenses and bigger image distance from the scintillator at the cost of light collection efficiency. If a high light collection efficiency is required for energy resolution, photomultipliers can be placed at the other sides of the scintillator for this purpose.\par

\begin{table}[tb]
\centering
\begin{tabular}{r|c|c}
\textbf{Slice} 			& \textbf{Resolution} & \textbf{Resolution}\\
\textbf{Binning S}	&  \textbf{from fit error} &  \textbf{from residuals} \\
										& \textrm{(s. Fig.\ref{res_fiter})} & \textrm{(s. Fig.\ref{res_line})} \\
\hline
1	& 68.8 \textmu m & - \\
2 & 45.0 \textmu m & 202.3 \textmu m\\
4 & - & 168.3 \textmu m \\
5 & 40.0 \textmu m & \\
8 & - & 136.6 \textmu m \\
10 & 38.3 \textmu m & - \\
\end{tabular}
\caption{The position resolution evaluated as the average fit error (row 2 and Fig. \ref{res_fiter}) and the average deviations from a straight line (row 3 and Fig. \ref{res_line}) multiplied with the pixel size (55 \textmu m) and divided by the magnification (3.3).}
\label{tab_res}
\end{table}

%
Some possible applications can be though of in the future: One of these could be the search for neutrinoless double beta decay. In this case high resolution particle tracking is a valuable tool to identify background events and enhance the significance of the observation. Another possible application could be high energy single photon Compton imaging where particle tracking can be used to determine the momentum direction of the Compton scattered electron.\par
An additional application could be beam profile monitoring at particle accelerators: Highly energetic partices excite the rest gas in the beam pipes which scintillates. An imaging of this beam profile from multiple directions could be useful for precise beam adjustments.

\section*{Acknowledgements}

We cordially thank the Medipix collaboration for the development of the HPD that we have used. This work was carried out within the Medipix collaboration.\par
We would like to thanks Ralf Diener and Samuel Ghazaryan from DESY for their support at the T-24 Testbeam line. Also, we would like to thank Dirk Wiedner from the MuPix Group for a share of their beam-time. We would like to thank Felix Just and Andrea Cavanna for their support. Also, we would like to thank Simon Filippov for proof-reading and valuable discussions.


\begin{thebibliography}{10}

\bibitem{bubble_chamber_rev}
W.B. Fretter, Ann. Rev. Nucl. Sci. \textbf{5} (1955), pp. 145-178

\bibitem{wire_chamb_rev}
C. Charpak and F. Sauli, Nucl. Instr. Meth. \textbf{162} (1979), pp. 405-428

\bibitem{tpc_review}
H. J. Hilke, Rep. Prog. Phys. \textbf{73} (2010), 116201

\bibitem{ssd_review}
L.M. Monta\~{n}o, J. Phys.: Conf. Ser. \textbf{18} (2005), 368

\bibitem{lhcb}
P. Collins et al., Nucl. Instr. and Meth. A \textbf{636} (2011), S185

\bibitem{lhcb_tp}
K. Akiba et al., Nucl. Instr. and Meth. A \textbf{661} (2012), pp. 31-49

\bibitem{medipix10}
M. Campbell, Nucl. Instr. and Meth. A \textbf{633} (2011), pp.1-10

\bibitem{dbreview}
S.M. Bilenky, C. Giunti, arXiv:1203.5250 (2012)

\bibitem{dmreview}
M. Schumann, arXiv:1310.5217v2 (2013)

\bibitem{tp_pub}
M. Filipenko, T. Gleixner et al., Eur. Phys. J. C \textbf{73} (2013), 73:2374

\bibitem{ahep}
T. Michel, T. Gleixner et al., AHEP \textbf{2013} (2013), 105318

\bibitem{next}
V. Álvarez et al., JINST \textbf{8} (2013), P09011

\bibitem{xenon}
E. Aprile and T. Doke, Rev. Mod. Phys. \textbf{82} (2010), pp.2053-2097

\bibitem{hpd_paper}
J. Vallerga, J. DeFazoi et al., submitted to iWoRid Conf. Proc. (2014)

\bibitem{fitpix}
M. Platkevic, J. Jakubek, Z. Vykydal et al., Nucl. Instr. Meth. A \textbf{591} (2012), pp. 254f

\bibitem{pixelman}
D. Turecek, J.Jakubek, Z. Vykydal et al., JINST \textbf{6} (2011), C01046

\bibitem{xavi}
X. Llopart, R. Ballabriga, M. Campbell, L. Tlustos and W. Wong, Nucl. Instr. Meth. A \textbf{581} (2007), pp. 485-494

\bibitem{blur_circle}
Harold M. Merklinger, Bedford, Nova Scotia: Seaboard Printing Limited (1993), ISBN 0-9695025-2-4

\bibitem{hough}
R.O. Duda and P.E. Hart, Comm. ACM \textbf{15} (1972), pp. 11-15

\end{thebibliography}
\end{document}